\documentstyle[12pt]{article}
\begin{document}

\title{THE ISRAEL THEOREM: WHAT IS NATURE TRYING TO TELL US?}
\author{L. Herrera\thanks{Postal address: Apartado 80793, Caracas 1080 A, Venezuela; e-mail address:
laherrera@cantv.net.ve}
\\
Escuela de F\'\i sica, Facultad de Ciencias,\\
Universidad Central de Venezuela,\\
Caracas, Venezuela.\\
}

\date{}
\maketitle

\begin{abstract}
We explore the possible physical consequences derived from the fact that the only static and asymptotically-flat vacuum space-time possessing a regular horizon is the Schwarzschild
solution (Israel theorem). If small deviations from the Schwarzschild metric are described by means of exact solutions to Einstein equations (as it should be), then for very compact
configurations, at the time scale at which radiatable multipole moments are radiated away, important physical phenomena should occur, as illustrated by  some
results on different solutions beloging to the Weyl class of static axially--symmetric solutions to the Einstein equations. 
\end{abstract}
\newpage
\section{Introduction}

It is a well established fact \cite{1}, that the only static and asymptotically-flat vacuum space-time possessing a regular horizon is the Schwarzschild solution. For all
the others Weyl exterior solutions \cite{2}, the physical components of the Riemann tensor exhibit singularities at $r=2M$. This result is usually refererred to as the Israel
theorem.

On the other hand, we know that all physical systems are submitted to fluctuations and, of course,  this  also applies to self--gravitating systems.
Accordingly we have to accept that any physical property of such a  system, e.g. spherical symmetry, is  to be submitted to such fluctuations.

Now, if the field produced by a self--gravitating system is not particularly intense (the boundary of the source is
much larger than the horizon) and fluctuations  off spherical symmetry are
sligth, then there is no problem in
representing the corresponding deviations from spherical symmetry (both inside and outside the source)
as a suitable perturbation of the spherically symmetric exact solution \cite{Letelier} (although strictly speaking the term ``horizon'' refers to the spherically symmetric case, we
shall use it when considering the $r=2M$ surface, in the case of small deviations from sphericity). 

However, as the object becomes more
and more compact, such perturbative scheme will eventually fail close to the source. Indeed, as is well known \cite{4}, though usually
overlooked, as the boundary surface of the source approaches the horizon (in the sense indicated above), any finite
perturbation of the Schwarzschild spacetime, becomes fundamentally different
from the corresponding exact solution representing the quasi--spherical spacetime,  even if the latter is characterized by parameters
whose values are arbitrarily close to those corresponding to Schwarzschild metric. This in turn is just an expression of the Israel theorem.

In other words, for strong gravitational
fields, no matter how small are the multipole moments (higher than monopole)
 of the source, there exists a bifurcation between the perturbed Schwarzschild metric  and all the
other Weyl metrics (in the case of gravitational perturbations). Examples of such a  bifurcation have been brought out in the study of the trajectories of test particles in the
$\gamma$ spacetime \cite{zipo}, and  in the $M-Q$ spacetime \cite{yo}, for orbits close to $2M$ \cite{HS},\cite{Herrera1}. 

To conciliate the abovementioned situation with the existence of a Schwarzschild black--hole, the black hole has no--hair theorem is invoked, according to which, in the process of
contraction all (radiatable) multipole moments are radiated away \cite{Price}. 

Nevertheless, the situation is more complex than it looks at first sight. Indeed, let us admit that in the process of collapse, all (radiatable) multipole moments are radiated away by
some, so far, unspecified mechanism. Obviously, such mechanism, as any physical process, must act at some time scale (say $\tau_{mech.}$). Now,  if it can be shown $\tau_{mech.}$ is
smaller than the time scale of any physical process occuring on the object (say
$\tau_{phys.}$), then the appearance of a horizon  proceeds safely.

However, let us suppose for a moment that there is a physical process whose  $\tau_{phys.}$
is of the order of magnitude of (or still worse, smaller than) $\tau_{mech.}$. In this case any physical experiment based on such process ``will see'' a
singularity as  the boundary of the object crosses the horizon, due to the always present fluctuations.

Indeed, the fact remains that  perturbations of spherical symmetry take place all along the evolution of the object. Thus, even if it is true that close to the horizon, any of these
perturbations is radiated away, it is likewise true that this is a continuous process. Then, as soon as a  ``hair'' is radiated away, a new  perturbation appears which will be later
radiated and so on. Therefore, since  ``hairs'' are radiated away at some {\it finite} time scale, then at that time scale ($\tau_{mech.}$) there will be always a  fluctuation acting
on the system.

Thus, unless one can  prove that indeed  $\tau_{mech.}$ is smaller than 
$\tau_{phys.}$ {\it{for any physical process}}, one should  take into account the possible consequences derived from the presence of fluctuations of spherical symmetry (close to
the horizon). 

Now, due to the bifurcation mentioned above, a fundamental question arises: How should we describe the quasi--spherical space--time resulting from the fluctuations off Schwarzschild?;
by means of a perturbed Schwarzschild metric or by means of an exact solution to Einstein equations, whose (radiatable) multipole moments are arbitrarily small, though non--vanishing?

Our point of view is that the description of such deviations should be done from  an exact solution of Einstein equations (of the Weyl family, if we restrict ourselves to vacuum static
axially--symmetric solutions) continuously linked to the Schwarzschild metric through one of its parameters, instead of considering a perturbation of the Schwarzschild space--time.

  This point of view is reinforced by some results obtained in the study of the source of quasi--spherical spacetimes
\cite{Herrera},
\cite{Herreramass}, where it has been shown that such bifurcation between the exactly spherically symmetric case and a solution of the Weyl family, when considering  the source (the
interior), takes place for strong gravitational fields, when the boundary surface is close to, but at a finite distance from, the horizon.

However before proceeding further with the discussion we should mention an important open question related to the propossed approach, namely: since there are as many
different (physically distinguishable \cite{luisyo}) Weyl solutions as there are different harmonic functions, then the  obvious  question arises: which among Weyl solutions is better
entitled to describe small deviations from spherical symmetry?.

In the past  \cite{HS}, \cite{Herrera1} , \cite{Herrera}, \cite{Herreramass} we have used the $\gamma$ metric and the $M-Q$ spacetime as examples of Weyl solutions, for describing
small deviations from spherical symmetry . The rationale for this choice was, in the first case, that the exterior $\gamma $-metric corresponds to a solution of the
Laplace equation (in cylindrical coordinates) with the same singularity
structure as the Schwarzschild solution (a line segment). In
this
sense the $\gamma $-metric appears as the ``natural'' generalization of
Schwarzschild space-time to the axisymmetric case. 

On the other hand, due to its relativistic multipole structure, the M--Q solution (more exactly, a sub--class of this solution
M-Q$^{(1)}$,
\cite{yo}) may be interpreted as a quadrupole correction to the Schwarzschild space--time, and therefore  represents a good candidate among known Weyl solutions, to describe small
deviations from spherical symmetry.

However it should be obvious that the question above has not a unique answer (there is an infinite number of ways of being non--spherical, so to speak) and therefore in the study of
any specific problem, the choice of the corresponding Weyl spacetime has to be reasoned.

All this having been said, let us now analyze the possible scenarios which might appear when the boundary surface of the object is sufficiently close to the horizon.
\vskip 1cm

\section{Scenarios}
\begin{itemize}

\item As a first scenario let us mention the standard point of view, according to which, in the process of collapse, all (radiatable) multipole moments  are radiated away leaving an
exactly spherically symmetric black hole, {\it and} it is assumed that $\tau_{mech.}$ is actually smaller than
$\tau_{phys.}$ for any physical process.

\item It is also possible that General Relativity would require some modifications for the case of  strong fields. It should be recalled that direct observational evidence supporting
this theory exists only for weak fields.

\item However, we believe that  General Relativity is valid for strong fields, and  nature is trying to tell us something through Israel theorem. Therefore whenever  $\tau_{mech.}$ is
equal to or larger  than
$\tau_{phys.}$ for some physical process, then as the the object becomes closer and closer to the horizon, important phenomena related to the presence of fluctuations off spherical
symmetry will apppear. The very nature of these phenomena depends on the specific Weyl solution representing the quasi--spherical spacetime. 
\end{itemize}

Within the context of this last scenario, let us recall that in the study of the
behaviour of some sources of the $\gamma$ metric, the inevitability of collapse in
the
spherical case, appears to be modified by a sharp increase in the
effective
inertial mass density term (or a sharp decrease in the "total force" term)
 as the surface gravitational potential approaches its maximum allowed value  \cite{Herrera}.
This increase makes the system more stable, hindering its departure from
equilibrium.

A similar conclusion was reached in \cite{Herreramass}, where it was shown that  the departure from equilibrium appears to be affected by a sharp modification  in the
stabilyzying  term of the active gravitational mass,
 as the surface gravitational potential approaches its maximum allowed value. 

Also, in the context of the $\gamma$ metric, it is worth noticing that the area surface of the source,  as it contracts and approaches the horizon, vanishes, thereby providing  a clue for the resolution of the information loss paradox \cite{paradox}.

Important differences with respect to the spherically symmetric case, also appear in the behaviour of
radial geodesics of test particles in the M-Q$^{(1)}$ metric \cite{Herrera1}. Thus, it has been established that particles along the axis of symmetry, close to the horizon, feel a
repulsive force for
an oblate source, and arbitrarily  small values of the quadrupole moment. 

Of course all these examples refer to specific Weyl metrics (and in some cases, to specific interiors of specific Weyl metrics \cite{sources}), but this just brings out the richness of
physical phenomena related to fluctuations of spherical symmmetry of compact objects in the context of general relativity, and  which might be of astrophysical relevance.

Finally it is  worth noting that we have referred exclusively to non--rotating sources. However we know that, on the one hand,  a result similar to Israel theorem exists for
stationary solutions with respect to the Kerr metric \cite{carter} and on the other, that almost all compact objects in nature are endowed with angular momentum. Accordingly, it should
be expected that the number and the richness of possible  astrophysical phenomena related to fluctuations off Kerr (described by means of exact stationary solutions to Einstein
equations) would  be substantially increased by the rotation of the source.


\begin{thebibliography}{88}

\bibitem{1} W. Israel, {\it Phys. Rev.} {\bf 164}, 1776 (1967).

\bibitem{2} H. Weyl,{\it Ann. Phys.} (Leipzig), {\bf 54}, 117 (1917);  H. Weyl,
{\it Ann. Phys.} (Leipzig), {\bf 59}, 185 (1919);  T. Levi.Civita,{\it Atti. Accad.
Naz. Lincei Rend. Classe
Sci.Fis. Mat. e Nat.}, {\bf 28}, 101 (1919);  J.L. Synge, {\it Relativity, the
general theory} (North-Holland Publ. Co, Amsterdam), (1960); D. Kramer, H. Stephani, M.A.H. MacCallum,  and E. Herlt,
{\it Exact Solutions of Einstein's Field Equations} (Cambridge University
Press, Cambridge) (1980).

\bibitem{Letelier} B.Boisseau, P.Letelier, {\it Gen.Rel.Grav.} {\bf 34},1077 (2002).

\bibitem{4} J.  Winicour, A.I. Janis and E.T. Newman, {\it  Phys. Rev.}
{\bf 176},1507 (1968); A.  Janis, E.T Newman and J. Winicour,{\it Phys.
Rev. Lett.} {\bf 20}, 878 (1968); F.I. Cooperstock and G.J. Junevicus {\it
Nuovo Cimento} {\bf
16B}, 387 (1973); L.  Bel, , {\it Gen. Relativ. Gravitation} {\bf 1}, 337 (1971).

\bibitem{zipo} R. Bach and H. Weyl, {\it Math. Z.}, {\bf 13}, 134 (1920); G.
Darmois, {\it Les equations de la Gravitation Einsteinienne} (Gauthier-Villars,
Paris) P.36, (1927);  D.M. Zipoy,{\it  J.
Math. Phys.}, {\bf 7}, 1137 (1966); R. Gautreau and J.L. Anderson, {\it Phys.
Lett.}, {\bf 25A}, 291 (1967); F.I. Cooperstock and G.J. Junevicus, {\it Int. J.
Theor. Phys.}, {\bf 9}, 59 (1968);
B.H. Vorhees,{\it  Phys. Rev.} D, {\bf 2}, 2119 (1970);F. Esp\'osito and L.
Witten, {\it Phys. Lett.}, {\bf 58B}, 357 (1975); K.S. Virbhadra, {\it Directional
naked singularity in General Relativity},
preprint gr-qc/9606004.

\bibitem{yo} J.L. Hern\'andez-Pastora and J. Mart\'\i n,{\it Gen.Rel.Grav.},
{\bf 26}, 877, (1994).

\bibitem{HS} L. Herrera, F. Paiva and N. O. Santos, {\it Int. J. Modern
Phys.D} {\bf 9}, 649 (2000).

\bibitem{Herrera1} L. Herrera, 2004, {\it Foun. Phys. Lett.} {\bf 18}, 21.

\bibitem{Price} R. Price, 1972, {\it Phys. Rev. D} {\bf 5}, 2419, 2439.

\bibitem{Herrera} L. Herrera, A. Di Prisco and J. Martinez, {\it Astr.
Space Sci.} {\bf 277}, 447 (2001).

\bibitem{Herreramass} L. Herrera, A. Di Prisco and E. Fuenmayor, {\it Class. Quantum Grav.} {\bf 20}, 1125 (2003).

\bibitem{luisyo} L. Herrera and J.L.  Hernandez--Pastora, {\it J. Math. Phys.} {\bf 41}, 7544,(2000).
\bibitem{paradox} L. Herrera {\it arXiv: 0709.4674}.
\bibitem{sources} R. Berezdivin,  B. Stewart, D. Papadopoulos, L. Witten and L. Herrera {\it Gen. Rel. Grav.} {\bf 14}, 97 (1982); L. Herrera, W. Barreto and J.L. Hernandez--Pastora
{\it Gen. Rel. Grav} {\bf 37}, 873, (2005); L. Herrera, G. Magli and D. Malafarina {\it Gen. Rel. Grav.} {\bf 37} 1371, (2005).

\bibitem{carter} B. Carter, {\it Phys. Rev. Lett.} {\bf 26} 331 (1971); S. W. Hawking, {\it Phys. Rev. Lett.} {\bf 26} 1344 (1971); R. Wald, {\it Phys. Rev. Lett.} {\bf 26} 1653 (1971).


\end{thebibliography}
\end{document}